\documentstyle[12pt,
epsfig]{article}
\begin{document}
\newcommand{\tcr}{T_{cr}}
\newcommand{\sxj}{\sigma_j}
\newcommand{\vt}{\tilde{v}}
\newcommand{\sxjt}{\tilde{\sigma_j}}
\newcommand{\dtt}{\tilde{\delta}}
\newcommand{\Lt}{\tilde{\Lambda}}
\newcommand{\Ut}{\tilde{U}}
\newcommand{\dj}{\delta j}
\newcommand{\du}{\delta u}
\newcommand{\Vt}{\tilde{V}}
\newcommand{\Ct}{\tilde{C}}
\newcommand{\dm}{\delta \mu}
\newcommand{\dms}{\delta m^2}
\newcommand{\dmst}{\delta \tilde{m}^2}
\newcommand{\dlx}{\delta \lambda}
\newcommand{\phit}{\tilde{\phi}}
\newcommand{\sxt}{\tilde{\sigma}}
\newcommand{\sxb}{\bar{\sigma}}
\newcommand{\lxb}{\bar{\lambda}}
\newcommand{\gb}{\bar{g}}
\newcommand{\rht}{\tilde{\rho}}
\newcommand{\jt}{\tilde{j}}
\newcommand{\djt}{\delta \tilde{j}}
\newcommand{\done}{\delta\phi_1}
\newcommand{\dy}{\delta y}
\newcommand{\tr}{\rm Tr}
\newcommand{\sx}{\sigma}
\newcommand{\mpl}{m_{Pl}}
\newcommand{\Mpl}{M_{Pl}}
\newcommand{\lx}{\lambda}
\newcommand{\Lx}{\Lambda}
\newcommand{\kx}{\kappa}
\newcommand{\ex}{\epsilon}
\newcommand{\be}{\begin{equation}}
\newcommand{\ee}{\end{equation}}
\newcommand{\eesn}{\end{subequations}}
\newcommand{\besn}{\begin{subequations}}
\newcommand{\beq}{\begin{eqalignno}}
\newcommand{\eeq}{\end{eqalignno}}
\def \lta {\mathrel{\vcenter
     {\hbox{$<$}\nointerlineskip\hbox{$\sim$}}}}
\def \gta {\mathrel{\vcenter
     {\hbox{$>$}\nointerlineskip\hbox{$\sim$}}}}

\newcommand{\nwc}{\newcommand}
%
%
\nwc{\cl}  {\clubsuit}
\nwc{\di}  {\diamondsuit}
\nwc{\sps} {\spadesuit}
\nwc{\hyp} {\hyphenation}
\nwc{\ba}  {\begin{array}}
\nwc{\ea}  {\end{array}}
\nwc{\bdm} {\begin{displaymath}}
\nwc{\edm} {\end{displaymath}}
\nwc{\bea} {\be\ba{rcl}}
\nwc{\eea} {\ea\ee}
\nwc{\ben} {\begin{eqnarray}}
\nwc{\een} {\end{eqnarray}}
\nwc{\bda} {\bdm\ba{lcl}}
\nwc{\eda} {\ea\edm}
\nwc{\bc}  {\begin{center}}
\nwc{\ec}  {\end{center}}
\nwc{\ds}  {\displaystyle}
\nwc{\bmat}{\left(\ba}
\nwc{\emat}{\ea\right)}
\nwc{\non} {\nonumber}
\nwc{\bib} {\bibitem}
\nwc{\lra} {\longrightarrow}
\nwc{\Llra}{\Longleftrightarrow}
\nwc{\ra}  {\rightarrow}
\nwc{\Ra}  {\Rightarrow}
\nwc{\lmt} {\longmapsto}
\nwc{\pa} {\partial}
\nwc{\iy}  {\infty}
\nwc{\ovl}  {\overline}
\nwc{\hm}  {\hspace{3mm}}
\nwc{\lf}  {\left}
\nwc{\ri}  {\right}
\nwc{\lm}  {\limits}
\nwc{\lb}  {\lbrack}
\nwc{\rb}  {\rbrack}
\nwc{\ov}  {\over}
\nwc{\pr}  {\prime}
\nwc{\nnn} {\nonumber \vspace{.2cm} \\ }
\nwc{\Sc}  {{\cal S}}
\nwc{\Lc}  {{\cal L}}
\nwc{\Rc}  {{\cal R}}
\nwc{\Dc}  {{\cal D}}
\nwc{\Oc}  {{\cal O}}
\nwc{\Cc}  {{\cal C}}
\nwc{\Pc}  {{\cal P}}
\nwc{\Mc}  {{\cal M}}
\nwc{\Ec}  {{\cal E}}
\nwc{\Fc}  {{\cal F}}
\nwc{\Hc}  {{\cal H}}
\nwc{\Kc}  {{\cal K}}
\nwc{\Xc}  {{\cal X}}
\nwc{\Gc}  {{\cal G}}
\nwc{\Zc}  {{\cal Z}}
\nwc{\Nc}  {{\cal N}}
\nwc{\fca} {{\cal f}}
\nwc{\xc}  {{\cal x}}
\nwc{\Ac}  {{\cal A}}
\nwc{\Bc}  {{\cal B}}
\nwc{\Uc}  {{\cal U}}
\nwc{\Vc}  {{\cal V}}
%
%
\nwc{\Th} {\Theta}
\nwc{\th} {\theta}
\nwc{\vth} {\vartheta}
\nwc{\eps}{\epsilon}
\nwc{\si} {\sigma}
\nwc{\Gm} {\Gamma}
\nwc{\gm} {\gamma}
\nwc{\bt} {\beta}
\nwc{\La} {\Lambda}
\nwc{\la} {\lambda}
\nwc{\om} {\omega}
\nwc{\Om} {\Omega}
\nwc{\dt} {\delta T}
\nwc{\Si} {\Sigma}
\nwc{\Dt} {\Delta}
\nwc{\al} {\alpha}
\nwc{\vph}{\varphi}
\nwc{\zt} {\zeta}
%
%
\def\tr{\mathop{\rm tr}}
\def\Tr{\mathop{\rm Tr}}
\def\Det{\mathop{\rm Det}}
\def\Im{\mathop{\rm Im}}
\def\Re{\mathop{\rm Re}}
\def\secder#1#2#3{{\partial^2 #1\over\partial #2 \partial #3}}
\def\bra#1{\left\langle #1\right|}
\def\ket#1{\left| #1\right\rangle}
\def\VEV#1{\left\langle #1\right\rangle}
\def\gdot#1{\rlap{$#1$}/}
\def\abs#1{\left| #1\right|}
\def\pr#1{#1^\prime}
\def\ltap{\raisebox{-.4ex}{\rlap{$\sim$}} \raisebox{.4ex}{$<$}}
\def\gtap{\raisebox{-.4ex}{\rlap{$\sim$}} \raisebox{.4ex}{$>$}}
\nwc{\Id}  {{\bf 1}}
\nwc{\diag} {{\rm diag}}
\nwc{\inv}  {{\rm inv}}
\nwc{\mod}  {{\rm mod}}
\nwc{\hal} {\frac{1}{2}}
\nwc{\tpi}  {2\pi i}
\def\contract{\makebox[1.2em][c]{
        \mbox{\rule{.6em}{.01truein}\rule{.01truein}{.6em}}}}
\def\slash#1{#1\!\!\!/\!\,\,}
%
%
\def\KK{{\rm I\kern -.2em  K}}
\def\NN{{\rm I\kern -.16em N}}
\def\RR{{\rm I\kern -.2em  R}}
\def\ZZ{Z \kern -.43em Z}
\def\QQ{{\rm \kern .25em
             \vrule height1.4ex depth-.12ex width.06em\kern-.31em Q}}
\def\CC{{\rm \kern .25em
             \vrule height1.4ex depth-.12ex width.06em\kern-.31em C}}
\def\ZZZ{Z\kern -0.31em Z}

\def \Msol {M_\odot}
\def\eV {\,{\rm  eV}}
\def\KeV {\,{\rm  KeV}}
\def\MeV {\,{\rm  MeV}}
\def\GeV {\,{\rm  GeV}}
\def\TeV {\,{\rm  TeV}}
\def\fm {\,{\rm  fm}}

\def\ap#1{Annals of Physics {\bf #1}}
\def\cmp#1{Comm. Math. Phys. {\bf #1}}
\def\hpa#1{Helv. Phys. Acta {\bf #1}}
\def\ijmpa#1{Int. J. Mod. Phys. {\bf A#1}}
\def\jpc#1{J. Phys. {\bf C#1}}
\def\mpla#1{Mod. Phys. Lett. {\bf A#1}}
\def\npa#1{Nucl. Phys. {\bf A#1}}
\def\npb#1{Nucl. Phys. {\bf B#1}}
\def\nc#1{Nuovo Cim. {\bf #1}}
\def\pha#1{Physica {\bf A#1}}
\def\pla#1{Phys. Lett. {\bf #1A}}
\def\plb#1{Phys. Lett. {\bf #1B}}
\def\pr#1{Phys. Rev. {\bf #1}}
\def\pra#1{Phys. Rev. {\bf A#1 }}
\def\prb#1{Phys. Rev. {\bf B#1 }}
\def\prp#1{Phys. Rep. {\bf #1}}
\def\prc#1{Phys. Rep. {\bf C#1}}
\def\prd#1{Phys. Rev. {\bf D#1 }}
\def\ptp#1{Progr. Theor. Phys. {\bf #1}}
\def\rmp#1{Rev. Mod. Phys. {\bf #1}}
\def\rnc#1{Riv. Nuo. Cim. {\bf #1}}
\def\zpc#1{Z. Phys. {\bf C#1}}
\def\APP#1{Acta Phys.~Pol.~{\bf #1}}
\def\AP#1{Annals of Physics~{\bf #1}}
\def\CMP#1{Comm. Math. Phys.~{\bf #1}}
\def\CNPP#1{Comm. Nucl. Part. Phys.~{\bf #1}}
\def\HPA#1{Helv. Phys. Acta~{\bf #1}}
\def\IJMP#1{Int. J. Mod. Phys.~{\bf #1}}
\def\JP#1{J. Phys.~{\bf #1}}
\def\MPL#1{Mod. Phys. Lett.~{\bf #1}}
\def\NP#1{Nucl. Phys.~{\bf #1}}
\def\NPPS#1{Nucl. Phys. Proc. Suppl.~{\bf #1}}
\def\NC#1{Nuovo Cim.~{\bf #1}}
\def\PH#1{Physica {\bf #1}}
\def\PL#1{Phys. Lett.~{\bf #1}}
\def\PR#1{Phys. Rev.~{\bf #1}}
\def\PRP#1{Phys. Rep.~{\bf #1}}
\def\PRL#1{Phys. Rev. Lett.~{\bf #1}}
\def\PNAS#1{Proc. Nat. Acad. Sc.~{\bf #1}}
\def\PTP#1{Progr. Theor. Phys.~{\bf #1}}
\def\RMP#1{Rev. Mod. Phys.~{\bf #1}}
\def\RNC#1{Riv. Nuo. Cim.~{\bf #1}}
\def\ZP#1{Z. Phys.~{\bf #1}}

\pagestyle{empty}
\noindent
\begin{flushright}
June 2004
\\
\end{flushright}
\vspace{3cm}
\begin{center}
{ \Large \bf
Energy from the Bulk through Parametric Resonance
\\
} \vspace{1.5cm} {\Large F. K. Diakonos, E. N. Saridakis and N.
Tetradis }
\\
\vspace{0.5cm}
{\it
Department of Physics, University of Athens,
Zographou 157 71, Greece
}
\\
\vspace{3cm}
\abstract{
We demonstrate that energy can be transferred very rapidly from the bulk
to the brane through parametric resonance. In a simple realization,
we consider a massless bulk field that interacts only with fields localized
on the brane. We assume an initial field configuration
that has the form of a wave-packet moving towards the brane. During the
reflection of the wave-packet by the brane the localized fields can be
excited through parametric resonance. The mechanism is also applicable to bulk
fields with a potential. The rapid energy transfer can have important
cosmological and astrophysical implications.
\\
\vspace{1cm}
PACS numbers: 11.10.Kk, 04.50, 98.80.Cq}
\end{center}

\newpage

\pagestyle{plain}

\setcounter{equation}{0}

\paragraph{Introduction:}

The idea that our Universe is a defect embedded in a higher-dimensional
bulk space is old \cite{rubakov}.
The Randall-Sundrum model \cite{rs} provides a realization of this idea.
The defect is a four-dimensional hypersurface
(a 3-brane) in a five-dimensional bulk with negative cosmological
constant (AdS space). The geometry is non-trivial (warped)
along the fourth spatial dimension, so that an effective localization of
low-energy gravity takes place near the brane. The matter is assumed to be
concentrated only on the brane.
For low matter
densities the cosmological evolution on the brane
becomes typical of a Friedmann-Robertson-Walker Universe \cite{binetruy,brax}.

Of particular interest in the scenario of a brane Universe
is the possibility of exchange of energy between
the defect and the bulk space. For this to occur
the bulk must contain some matter component in addition to the negative
cosmological constant.
The presence of matter in the bulk affects the expansion on the brane.
For an observer on the brane the
modification can be attributed to ``mirage'' matter components
\cite{mirage}.
The energy exchange can modify the cosmological evolution
as well \cite{vandebruck}. In particular,
the energy transfer from the bulk to the brane can lead to periods of
accelerated expansion on the brane \cite{tet}.

In this letter we concentrate on the mechanism that can lead to
a rapid energy transfer from the bulk to the brane. We omit the
gravitational effects and study a simplified system that consists of a
flat five-dimensional bulk space with a four-dimensional brane as its
boundary. In the bulk we assume the presence of a field that interacts
only with fields localized on the brane.
Our aim is to demonstrate that
energy can be transferred very efficiently from the bulk to the brane fields
through a process that resembles very strongly the parametric resonance
\cite{KLS,kofman}.
The incorporation of gravitational effects is technically complicated and
will be left for a future publication.

\paragraph{The model:}

We consider a massless bulk field $\phi$ that interacts with a field
$\chi$ localized on
the brane. We neglect gravitational effects and
assume that the
system is described by the action
\be
S= \int d^4x\,dy \left[
\partial^M \phi\, \partial_M \phi
+\delta(y) \left( \partial^\mu \chi \, \partial_\mu \chi
-V(\chi,\phi) \right) \right],
\label{lagr} \ee
where $M=0,1,2,3,4$ and $\mu=0,1,2,3.$ In order to keep our discussion
simple we identify the half-space $y<0$ with the half-space $y>0$, in
complete analogy to ref. \cite{rs}. In this way the brane forms the
boundary of the bulk space.
We assume that the potential $V(\chi,\phi)$
has the form
\be
V(\chi,\phi)=\frac{1}{2}m^2\chi^2 + \frac{1}{2}g\chi^2\phi^2.
\label{pot} \ee

We expand the quantum field $\chi$ as
\be
\chi(t,\vec{x})=\int \frac{d^3k}{(2\pi)^3}
\left(
a_k f_k(t)e^{-i\vec{k} \vec{x}}
+a_k^\dagger f_k^*(t)e^{i\vec{k} \vec{x}}
\right),
\label{chiex} \ee
where the mode functions $f_k$ satisfy $f_k\dot{f}_k^*-f_k^*\dot{f}_k=i$, and
the number distribution function is $n_k\equiv \langle a_k^\dagger a_k
\rangle$.
The equation of motion of $\chi$ gives
\be
\ddot{f}_k(t)+\left(k^2+m^2+g\phi^2(t,0)\right)f_k(t)=0,
\label{fk} \ee
where a dot denotes a time derivative.
We also have
\be
\langle \chi^2 \rangle (t)=\int\frac{d^3k}{(2\pi)^3}f^*_k(t)f_k(t)(1+2n_k).
\label{chiav} \ee

We treat the field $\phi$ classically.
We also make the simplifying assumption that the configuration of
$\phi$ depends only on the time $t$ and the fourth spatial
coordinate $y$. For this assumption to be consistent with the
$\phi$-field evolution
we replace $\chi^2(t,\vec{x})$ with $\langle \chi^2 \rangle (t)$
in the equation of motion of $\phi$. This is the
Hartree approximation \cite{cooper} that is usually employed in the
study of parametric resonance \cite{kofman}.
The $\phi$-field equation of motion becomes
\be
\ddot{\phi}(t,y)-\phi''(t,y)+ g \, \delta(y)\, \phi(y,t)\,
\langle \chi^2 \rangle(t)=0,
\label{phieq} \ee
where a prime denotes a $y$-derivative.
Integrating this equation with respect to $y$ in the interval $[-\ex,\ex]$ and
taking $\ex \to 0$, we find
\be
\phi'(t,0^+)=-\phi'(t,0^-)=\frac{1}{2}g \, \phi(t,0) \,
\langle \chi^2 \rangle(t).
\label{boundary} \ee
The solution of eq. (\ref{phieq}) is equivalent to the solution of
the bulk equation
\be
\ddot{\phi}(t,y)-\phi''(t,y)=0
\label{phieqb} \ee
with the boundary condition (\ref{boundary}).

In the following we solve numerically the system of equations
(\ref{fk}), (\ref{chiav}), (\ref{boundary}), (\ref{phieqb}) in the
region $y\geq 0$. The solution for negative $y$ is $\phi(t,-y)=\phi(t,y)$.
The initial configuration $\phi(0,y)$ is arbitrary.
We choose it to have
the form of a wave-packet that moves towards the brane. For
the mode functions
of the $\chi$-field for $t\to 0$, when $\phi(t,0)\simeq 0$,
we assume the vacuum expressions
$f_k(t)={e^{-i\omega_k t}}/{\sqrt{2\omega_k}}$,
with $\omega_k=\sqrt{k^2+m^2}$.
This gives $f_k(0)=1/\sqrt{2\omega_k}$,
$\dot{f}_k(0)=-i\omega_k/\sqrt{2\omega_k}$.
The evolution of $f_k(t)$ can be attributed to the creation of
$\chi$-particles on the brane through the time-varying $\phi$-field.
The time-dependent number distribution function
can be calculated through a Bogoliubov transformation \cite{cooper}.
It is approximately
\be
n_k(t)=\frac{1}{2}\omega_k(t)\left(
\left| f_k(t)\right|^2+\frac{\left| \dot{f}_k(t)\right|^2}{\omega_k^2(t)}
\right)(2n_k+1)-\frac{1}{2},
\label{number} \ee
with $\omega_k(t)=\sqrt{k^2+m^2+g\phi^2(t,0)}$.
The above expression has corrections $\sim \dot{\omega}_k$ that we
have neglected by assuming that $\dot{\omega}_k(t)\ll \omega^2_k(t)$.
In the limit $t \to \infty $, after the $\phi$-wave-packet has
been reflected by the brane and $\phi(t,0)\simeq 0$, our approximation
becomes exact. This means that the final number of $\chi$-particles is
computed exactly, even though the expression (\ref{number}) is approximate
when $\phi(t,0) \not= 0$.
The initial number distribution function $n_k$ can be chosen according to
our assumptions for the initial state of the physical system. For example,
it can have the form of a thermal distribution \cite{cooper}.
We make the simpler assumption that initially there is no significant
number density of $\chi$-particles, so that $n_k=0$.

Several approximations that we have made in deriving the evolution
equations for the fields are not expected to affect the qualitative
physical behaviour. For example, a term $\sim \chi^4$ in
the potential of eq. (\ref{pot}) has been taken into account in previous
studies of parametric resonance, without altering significantly the
essential behaviour.
The Hartree approximation becomes exact in the large-$N$
limit, if the field $\chi$ is assumed to have $N$ components and the
Lagrangian to be invariant under an $O(N)$ symmetry.
Finally, the possible presence
of a potential for the bulk field is not essential
either. As we shall see in the following, the necessary requirement
is that the bulk field must be oscillatory at the location of the brane.
Any wave solution that is reflected by the brane is capable of transferring
energy onto it through parametric resonance.

\paragraph{An analytical solution:}

In order to have some intuitive understanding of the solutions of
interest we derive in this section an analytical solution of
eqs. (\ref{fk}), (\ref{chiav}), (\ref{boundary}), (\ref{phieqb}).
We make the assumption $\phi(t,0)=\phi_0\cos\alpha t$.
Then, eq. (\ref{fk}) can be put in the form of the Mathieu equation
\be
\frac{d^2f_k}{dz^2}+(A_k-2q\cos 2z)f_k=0,
\label{mathieu} \ee
with
$z=\alpha t-\pi/2$, $A_k=(k^2+m^2+g\phi_0^2/2)/\alpha^2$,
$q=g\phi_0^2/2\alpha^2$.
An infinite number of resonance bands exist in $k$, within which the amplitude
of $f_k$ grows exponentially fast with time. As a result, $\langle \chi^2
\rangle (t)$ grows fast.

The form of $\phi(t,y)$ that is consistent with the above solution for
$\chi$ can be determined analytically. It is given by the solution of the
wave equation (\ref{phieqb}) for $y\geq 0$,
with the ``initial'' conditions given for $y=0$. Essentially the role of
$t$ and $y$ is reversed compared to the standard case.
The most general solution of
eq. (\ref{phieqb}) is
\be
\phi(t,y) = \frac{1}{2}\left[
a(t+y)+a(t-y)+\int_{t-y}^{t+y}b(t')dt'
\right],
\label{ansol1} \ee
where the functions $a(t)$, $b(t)$ are determined through the boundary
conditions at $y=0$. In particular,
$a(t)=\phi(t,0)=\phi_0\cos\alpha t$ according to our assumption.
The function $b(t)$ is defined through eq. (\ref{boundary}), that gives
$b(t)=\phi'(t,0)=g\phi(t,0)\langle \chi^2 \rangle(t)/2$.

For a given time $t$ the field $\phi(t,y)$
has an oscillatory form with an envelope that grows
as a function of $y$. As time increases the slope of
the envelope grows. This behaviour is a consequence of our assumption
that $\phi$ has a constant amplitude $\phi_0$ at $y=0$. The growth
of $\langle \chi^2 \rangle(t)$ induces a backreaction on the evolution of
$\phi$ that tends to reduce its amplitude. This effect can be compensated
only if the amplitude of the bulk oscillations of $\phi$ increases with
time.

\begin{figure}[t]
 \begin{center}
 \mbox{\epsfig{figure=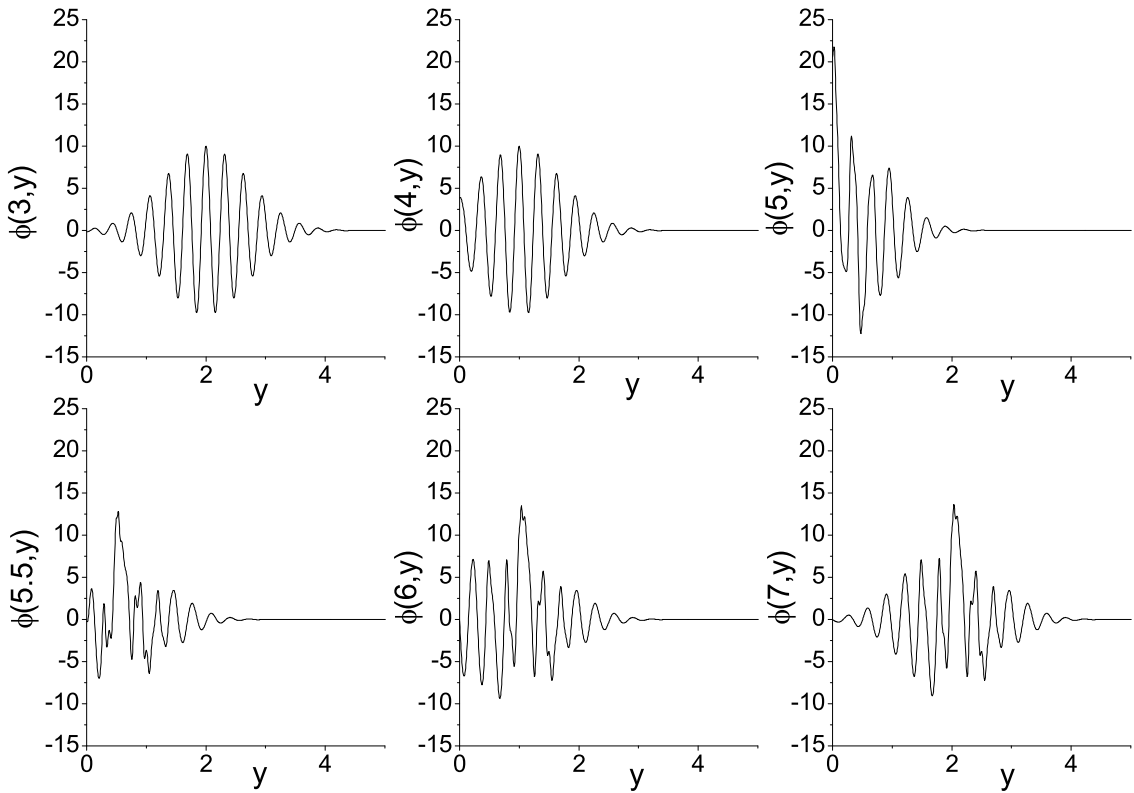,width=17cm,angle=0}}
 \caption{\it
The form of the bulk field at various times.
 }
 \label{fig1}
 \end{center}
 \end{figure}

 \begin{figure}[t]
 \begin{center}
 \mbox{\epsfig{figure=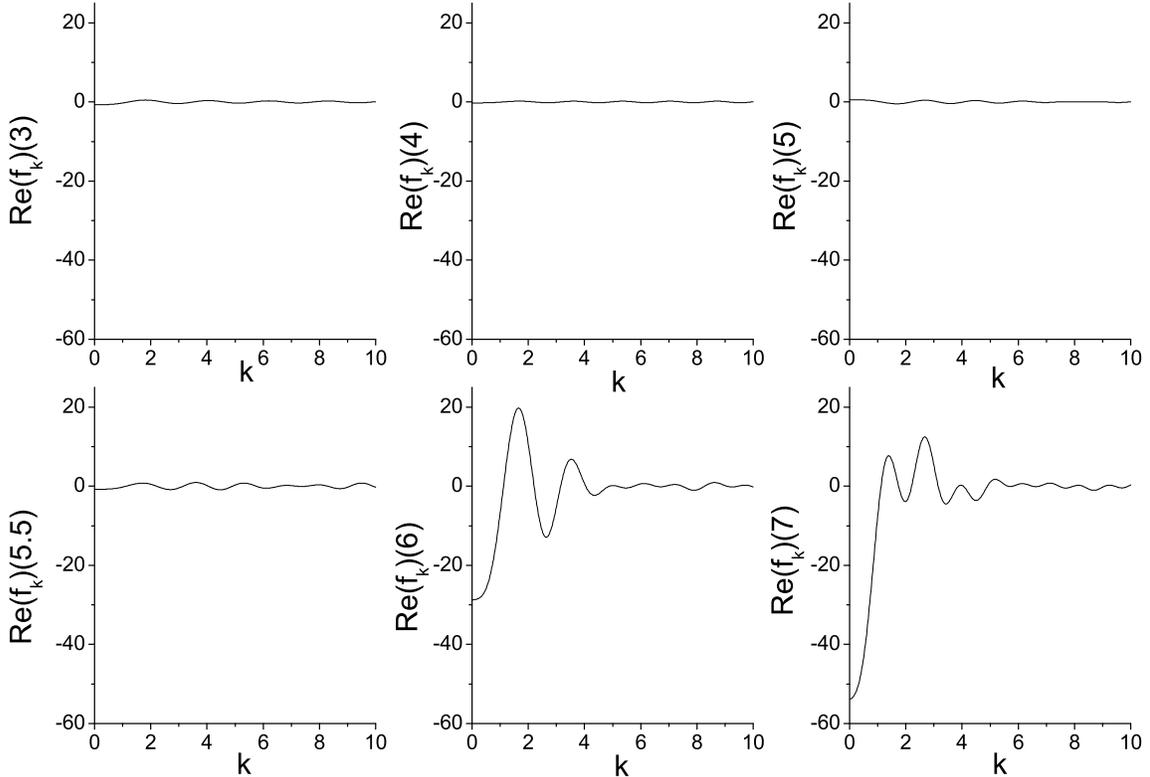,width=17cm,angle=0}}
 \caption{\it
The real part of the mode function $f_k$ of the brane field at various times.
 }
 \label{fig2}
 \end{center}
 \end{figure}

 \begin{figure}[t]
 \begin{center}
 \mbox{\epsfig{figure=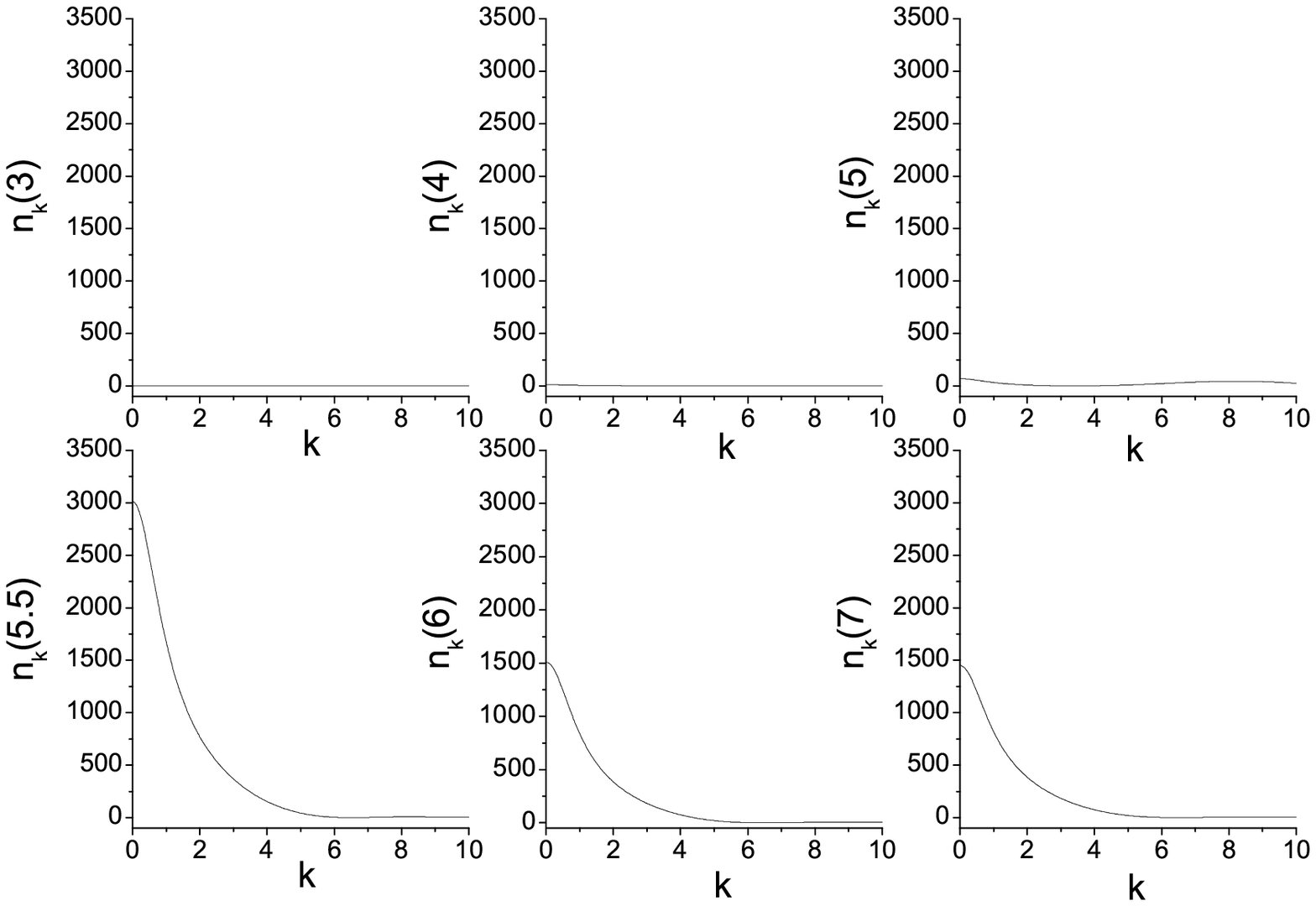,width=17cm,angle=0}}
 \caption{\it
The number distribution function $n_k$ on the brane at various times.
 }
 \label{fig3}
 \end{center}
 \end{figure}

\paragraph{The numerical solution:}

The analytical solution indicates that the
rapid transfer of energy from the bulk to the brane through parametric
resonance is possible. However, the solution was constructed by
assuming the
form of the $\phi$-field at the location of the brane and deducing its form
in the bulk. Here we would like to start by defining initial
conditions for the system, such that the deviation of $\phi$ from zero is
significant only
away from the brane. Subsequently, the $\phi$-field perturbation
moves close to the brane and leads to the excitation of the $\chi$ field.
This calculation cannot be performed analytically because of the difficulty
in implementing the boundary condition (\ref{boundary}). For this reason
we solve the system of equations
(\ref{fk}), (\ref{chiav}), (\ref{boundary}), (\ref{phieqb}) numerically.

For early times
we consider a configuration for the $\phi$-field
given by
\be
\phi(t,y)=\phi_0\exp \left(
-\frac{(y-y_0+t)^2}{\sx^2}\right)\cos\left[\alpha (y-y_0+t)\right].
\label{init1} \ee
It corresponds to a wave-packet of fundamental frequency $\alpha$
and width $\sx$, centered at
$y=y_0-t$, that moves towards the
brane located at $y=0$ with the speed of light.
In the following we study the evolution of a wave-packet with
$\phi_0=10$, $y_0=5$, $\sx=1$, $\alpha=20$.
We use the values
$m^2=1$, $g=10$ for the parameters in the potential of
eq.(\ref{pot}). Essentially, we express all dimensionful quantities in
units of $m$.
We also assume a negligible initial particle number density:
$n_k=0$. The total number density of $\chi$-particles and their energy
density are given by
\be
\frac{N(t)}{V}=\int \frac{d^3k}{(2\pi)^3} n_k(k),
~~~~~~~~~~~~~~
\frac{E(t)}{V}=\int \frac{d^3k}{(2\pi)^3} \omega_k(t)\, n_k(k),
\label{ennum} \ee
with $n_k(t)$ given by eq. (\ref{number}).
The momentum integrations in these equations, as well as in
eq. (\ref{chiav}),
have ultraviolet divergences. We regulate them through an explicit cutoff
$\Lx=10$. Moreover, the time-evolution of the $\chi$-field is
expected to display the
well-known ultraviolet problem of classical systems. (The Hartree approximation
in the leading order leads to essentially classical evolution.)
The high-frequency modes become excited at sufficiently large times
and the ultraviolet contributions dominate in
the expressions for the various observables.
In our simplified study we omit all
$\chi$-field self-interaction terms. As a result, the ultraviolet modes can
be excited only during the finite interval that the wave-packet interacts
with the brane. For our choice of parameters, no significant excitation of
these modes takes place.

In fig. 1 we display the evolution of the wave-packet. At $t=0$ it is
located away from the brane. Subsequently it moves towards the brane and
interacts with it during the time interval $3<t<7$.
Eventually it gets reflected with a distorted profile. At times $t\simeq 5$
the maximum of $\phi(t,0)$ is $\simeq 2 \phi_0$.
The reason is that, because of our assumption of reflection symmetry around
$y=0$, there is another wave-packet interacting with the brane from the left.

 \begin{figure}[ht]
 \begin{center}
 \mbox{\epsfig{figure=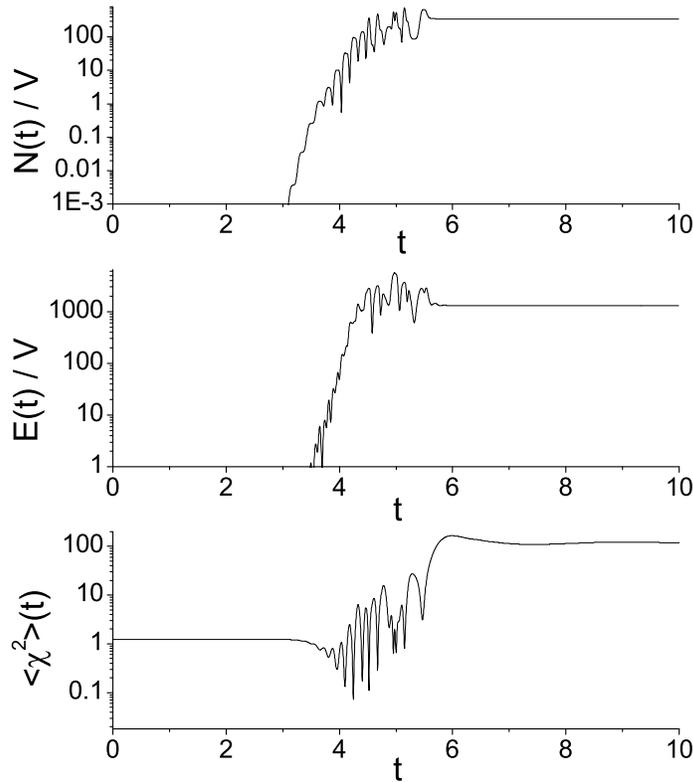,width=11.7cm,angle=0}}
 \caption{\it
The total number density of $\chi$-particles,
the energy density carried by them and the dispersion of the $\chi$-field,
as a function of time.
 }
 \label{fig4}
 \end{center}
 \end{figure}

In fig. 2 we depict the evolution of the mode functions $f_k(t)$ of the
$\chi$-field. We observe that they increase strongly in the region of small
$k$ through parametric resonance. The mode functions with $k$ near the cutoff
$\Lx=10$ are not excited significantly during the time interval that
we consider. This implies that we can trust the predicted physical
behaviour.

In fig. 3 we display the number distribution function $n_k(t)$ as given by eq.
(\ref{number}). We observe that during the time interval $3<t<7$ the
low-frequency modes are excited. The quantity $n_k(t)$ displays
some oscillatory behaviour with time, but on the average it grows fast.
For $t>7$ the wave-packet moves away from the brane and the distribution
function becomes constant. This is a consequence of the omission of
any $\chi$-field self-interactions in our simplified treatment.

In fig. 4 we plot the total number density of $\chi$-particles,
the energy density carried by them and the dispersion of the $\chi$-field.
All these quantities display
a rapid increase in the time interval $3<t<7$, while they remain constant
at later times when the wave-packet has moved away from the brane.

The intensity of the parametric resonance depends on the parameters of the
model. We can use the Mathieu equation (\ref{mathieu})
in order to obtain some intuition.
The amplification is
more pronounced if the system is in the region of broad
parametric resonance. This requires $g\phi^2_0/m^2 \gta 1$, so that
$A_k\gta q$. The parameters we chose satisfy this constraint.
The time dependence of $n_k(t)$ in fig. 3 is typical of
a system in the region of broad parametric resonance \cite{kofman}.

The resonance effects could be further enhanced
if the effective mass became zero for a certain value of $\chi$. This would
require $m^2<0$ and would imply that
the $\phi$-field determines the $\chi$ expectation value. We have used
a simpler model, in which the bulk plays only a minor role in the determination
of the vacuum state of the brane fields, while it can affect their dynamical
evolution.

Despite the similarity we described above, a careful analysis indicates that
the Mathieu equation is not strictly relevant if the bulk field has
the form of a wave-packet. The reason is that
the function $\phi(t,0)$ that appears in eq. (\ref{fk})
describes an oscillation with an amplitude that is non-zero
only  within a finite time interval. The Fourier transform of this function
is non-zero over a large range of frequencies, and not for a unique
frequency as in the Mathieu equation. We expect a qualitative behaviour similar
to the one predicted by the Lam\'e equation.
In agreement with this conclusion, we do not
find any parameter range that resembles the region of narrow
parametric resonance. The numerical analysis of the range
$g\phi^2_0/m^2 \lta 1$ leads to an evolution very similar to the one
depicted in figs. 1--4. The only difference is that the number density of
the produced $\chi$-particles and the total energy trasferred to the brane
fall very fast with decreasing $g\phi^2_0/m^2$.
When $g\phi^2_0/m^2 \ll 1$ the resonance effects become negligible.
Similarly, significant excitation of the $\chi$-modes is possible only if the
fundamental frequency $\alpha$
of the $\phi$ wave-packet is not much smaller than
the mass $m$.

\paragraph{Cosmological and astrophysical implications:}

The rapid transfer of energy from the bulk to the brane can affect the
cosmological evolution. If the $\chi$-field is coupled to lighter brane
particles the energy can be transferred to them through the decay of
the $\chi$-modes. This would result in a period like
preheating \cite{KLS,kofman}.
The two mechanisms are so similar
that the process we have described could be characterized as
preheating by a bulk source. There are several advantageous aspects
for this type of preheating:\\
a) The form of the source is less constrained than in the conventional
scenario. In particular,
the field $\phi$ need not be homogeneous. In the example we
discussed it has the form of a wave-packet that depends only on the
extra spatial coordinate. There is no need for a specially chosen potential
for $\phi$. In our example the potential was assumed to be zero. \\
b) If $\phi$ is not the field responsible for the brane inflation,
the time of preheating and the amount of the released energy
are not constrained. In our example, for $g\phi^2_0/m^2\gta 1$ the energy
density in the $\chi$-modes exceeds $m^4$. For large $m$ significant
amounts of energy can be deposited on the brane.

On the other hand, this scenario requires an additional
non-compact spatial dimension.
A picture consistent with observations
can be obtained only by
generalizing the Randall-Sundrum model \cite{rs} in a cosmological context.
The graviton localization near the brane leads to a low-energy
cosmological evolution typical of a four-dimensional Universe
\cite{binetruy,brax}.
The effect of the energy influx on the cosmological evolution can be more
complicated than the mere increase
of the brane energy density. It is possible that an era
of cosmological acceleration can take place \cite{tet}.
A detailed investigation of this issue must include the solution of
the Einstein equations for the generalized warped geometry
of the Randall-Sundrum model \cite{brax}.
A numerical study of such a system is technically difficult.
The only existing study has considered a two-brane
scenario without brane fields \cite{martin}.
The numerical investigation of the scenario with brane
fields is in progress.

Another manifestation of the mechanism we described could be related to the
energy spectrum of high energy cosmic rays. It has been argued that
these could be produced in the
decays of long-lived heavy particles of cosmological origin
\cite{cosmic,cosmic2}.
Our mechanism provides an alternative possibility: The decaying heavy
particles could correspond to the $\chi$-modes that are produced during the
energy transfer from the bulk to the brane. In order to provide an
explanation for the highest energy cosmic rays, the mass
of the $\chi$-particles must be $m\gta 10^{13}$ GeV \cite{cosmic,cosmic2}.
The production of so heavy particles is possible only if the
fundamental frequency $\alpha$ of the wave-packet is of the same
order of magnitude.

The total amount of transferred energy is controlled
by the combination $g\phi_0$, where $g$ is the strength of the coupling
between the bulk and brane fields and $\phi_0$
the maximum amplitude of the wave-packet.
The total amount of energy stored in
$\chi$-particles drops very fast for
$g\phi_0\ll m$, with $m$ the mass of the $\chi$-field.
In this range
the resonance effects become negligible.
Moreover, the dilution of the energy density through the possible accelerated
expansion on the brane because of the energy influx has not been taken
into account in our discussion.
The combination of the two factors indicates that the total amount of
transferred energy can be much smaller than $m^4$,
so as not to disturb the conventional
cosmological picture, even though the energy will be distributed in
heavy particles of mass $m$. The quantitative analysis of these issues
will be possible only through the numerical solution of the Einstein equations.

As a final comment we mention that the role of the bulk field can be
played by the gravitational sector of the theory. It is possible
that a  perturbation of the bulk metric that has the
form of a gravitational wave interacting with the brane can result in the
excitation of heavy brane particles. This is an effect opposite to
the production of bulk gravitons by the brane matter \cite{giudice}.

\paragraph{Acknowledgements:}
The work of E. N. Saridakis was supported by the Greek State
Scholarship's Foundation (IKY). The work of N. Tetradis was
partially supported through the RTN contract HPRN--CT--2000--00148
of the European Union. The authors acknowledge financial support
through the research programs ``Pythagoras'' and ``Kapodistrias''.

\end{document}